\documentclass[aps,prb,twocolumn]{revtex4}

\newcommand{\be}{\begin{equation}}
\newcommand{\ee}{\end{equation}}

\usepackage{graphicx}
\usepackage{amsmath}
\graphicspath{{.}{./EPS/}}

\begin{document}

\title{Statistical model of dephasing in mesoscopic
devices introduced in the scattering matrix formalism}

\author{Marco G.~Pala$^1$}
\author{Giuseppe Iannaccone$^{1,2}$}
\affiliation{$^1$Dipartimento di Ingegneria dell'Informazione,
Universit\`a degli Studi di Pisa,
via Caruso, I-56122 Pisa, Italy}
\affiliation{$^2$IEIIT-Consiglio Nazionale delle Ricerche, via Caruso,
I-56122 Pisa, Italy}
\date{\today}

\begin{abstract}
We propose a phenomenological model of dephasing
in mesoscopic transport, based on the introduction of random phase
fluctuations in the computation of the scattering matrix of the
system. A Monte Carlo averaging procedure allows us to extract
electrical and microscopic device properties. We show that, in
this picture, scattering matrix properties enforced by current
conservation and time reversal invariance still hold. In
order to assess the validity of the proposed approach, we present
simulations of conductance and magnetoconductance of Aharonov-Bohm
rings that reproduce the behavior observed in experiments,
in particular as far as aspects related to decoherence are concerned.
\end{abstract}


\maketitle

\section{Introduction}

Phase coherence of the electron wave function has a
fundamental influence on the transport properties of
mesoscopic devices \cite{stone} and is at the basis of
several phenomena ranging from interference effects (such as
Aharonov-Bohm oscillations) to weak localization (WL) \cite{abrahams,aronov}
and universal conductance fluctuations (UCF).\cite{altshuler,lee}

The fundamental quantity used to express the degree of coherence
in a system is the so-called phase coherence length 
(or ``dephasing'' length) $l_\phi$,\cite{been} 
that is typically estimated on the
basis of WL experiments in semiconductor heterostructures,
\cite{lin} Si MOSFETs \cite{bishop}, metal conductors,
\cite{mohanty} or of interference experiments in devices such as
Aharonov-Bohm (AB) rings.\cite{hansen,pedersen}

Mesoscopic physics deals with devices whose size is smaller or
comparable to $l_\phi$ and therefore often operate in an
intermediate regime between coherent transport, in which the phase
information is fully preserved, and incoherent transport. The main
phase-breaking mechanisms are due to interaction of electrons with
other electrons, photons, phonons, and defects such as magnetic
impurities or to other kinds of phase-randomizing interaction
with the environment. \cite{stern,chakravarty}

Therefore, it would be very useful to have a unique formalism
capable to include an arbitrary degree of dephasing in the evaluation
of the transport properties of a system, and to allow a seamless
transition between the coherent and the fully incoherent limits.

In the case of interfering paths, an ``ad-hoc" random term can be
added analytically to the difference between the phases accumulated
in the two paths. When generic devices with two or
more leads are considered, two main phenomenological models are
available for including a partial degree of dephasing in the
transport model: $i)$ insertion in the device of an additional
``virtual'' voltage probe \cite{buttiker} that can be also taken
into account by properly adjusting the two-terminal
conductance\cite{brouwer,beenakker}; or $ii)$ addition of an imaginary
part to the Hamiltonian in the device region.
\cite{czycholl,efetov,mccann}
In case $i)$, the seeming drawback of spatially localized decoherence
can be overcome either by introducing an adequate number of virtual probes
in different points of the device region,\cite{ando} 
or by considering the limit of a voltage lead that supports
an infinite number of modes.\cite{brouwer}
In case $ii)$, the carriers absorbed by the imaginary term
have to be re-injected into the conductor in order to 
ensure current conservation.

An additional method to treat dephasing consists in including a
stochastic absorption in the scattering description
\cite{benjamin,joshi,ianna} through the insertion of an attenuation
factor in the free propagation region. Also in this case,
continuity of the probability density current requires that
absorbed electrons are re-injected.

Dephasing due to the environment can be modeled by two equivalent
approaches:\cite{stern} 
One focuses on the changes that the wave function induces 
on the state of the environment, and was adopted, for instance,
to simulate electron conduction interacting with dynamic impurities.
\cite{mello}
The other addresses the phase accumulated by the interfering waves
as a statistical process.
In this paper, we adopt the latter perspective 
and propose a phenomenological model of
decoherence, that treats dephasing as a distributed phenomenon in
the device region, ensures the conservation of current density
and allows to evaluate the local density of states.
We consider the stochastic behavior of the dephasing process and
adopt a Monte Carlo averaging procedure to extract the electrical
and microscopic properties of the system.
We are able to vary $l_\phi$ and gradually move from a
coherent to a totally incoherent transport regime.
The model is described in Sec.~\ref{model} 
and is applied in Sec.~\ref{results}
to evaluate the decoherence on the conductance and
magnetoconductance of an AB ring.


\section{Dephasing model}
\label{model}

We include our model for dephasing in the scattering matrix
formalism for the computation of the device conductance $G$. The
conductance of a generic structure is related to the transmission
probability matrix $T=t^\dag t$ by the Landauer-B\"uttiker formula
$G=g e^2/h \, \sum_{n,m} \, T_{nm}$, \cite{landauer} where $t$ is
the transmission matrix, $g$ is the spin degeneracy factor, $e$ is
the elementary charge, $h$ is Planck's constant and $n$, $m$
run over all transverse modes contributing to transport.

The transmission matrix is obtained by computing the scattering
matrix (S-matrix) of the device.\cite{datta} If we
subdivide the domain along the transport direction $x$ in several
slices, one for each grid-point $j = 1,\dots,N_x$ in the $x$
direction, the wave function of the electron in the $j$-th slice
$(x_{j} < x < x_{j+1})$
can be written as
\be
\label{psi}
\psi_j(x,y)= \sum_n
\frac{\chi_{j,n}(y)}{\sqrt{|k_{j,n}|}} \left(a_{j,n} e^{i k_{j,n}
x}+b_{j,n} e^{-i k_{j,n} x} \right) \; ,
\ee
where $\chi_{j,n}(y)$ is the $n$-th transverse
eigenvector of the $j$-th slice with eigenenergies $E_{j,n}$
and the longitudinal wave vector $k_{j,n}$ is related to the total
energy $E$ by the condition $E=E_{j,n}+\hbar^2 k_{j,n}^2 /2m_j$.
The coefficients $a_{j,n}$ and $b_{j,n}$ are obtained by imposing
the continuity of the wave function and of the probability current density
at the interface between the $j$-th and the $(j+1)$-th slice. The
scattering matrix $S_j$ links the incoming and the outgoing
coefficients:
\be \left(
\begin{array}{c} b_{j}
\\ a_{j+1} \end{array} \right)= S_j \, \left( \begin{array}{c} a_{j}
\\ b_{j+1} \end{array} \right) \;,
\ee
where $a_j$ ($b_j$) is the column vector of all $a_{j,n}$'s ($b_{j,n}$'s)
for $n=1,\dots,N_{\rm mode}$,
and $N_{\rm mode}$ is the total number of modes considered in the system.
The composition between two adjacent scattering matrices\cite{datta}, $S_j$ and
$S_{j+1}$, gives the matrix $S_j \otimes S_{j+1}$ which links
$a_j$, $b_j$, $a_{j+2}$, and $b_{j+2}$.
In order to compute the scattering matrix of the complete device
we have to compose the matrices of all slices
according to well known rules:\cite{datta}
$S=S_1 \otimes \cdots \otimes S_j \otimes \cdots \otimes S_{N_x-1}$.

We model the effect of decoherence as a random variation of
the phase accumulated by each mode in each slice into which
the device has been divided. In the absence of dephasing,
mode $n$ accumulates in slice $j$ a phase
$k_{j,n}(x_{j+1}-x_j)$; in the presence of dephasing
it accumulates a phase $k_{j,n}(x_{j+1}-x_j) + \Delta \phi_{j,n}$,
where $\Delta \phi_{j,n}$ is a random term obeying a Gaussian
probability distribution with zero average and
standard deviation $\sigma_j$, that depends on the thickness
of the slice $(x_{j+1}-x_j)$ and on $l_\phi$, as we shall show.

For a random choice of all $\Delta \phi_{j,n}$'s,
for $j=1,\dots,N_x-1$, $n=1,\dots,N_{\rm mode}$, we
can compute an ``occurrence'' $\tilde{S}$ 
of the scattering matrix of the system.
We take into account the probabilistic nature of
dephasing, and therefore transport properties are
obtained following a Monte Carlo averaging procedure
over a sufficiently
large ensemble of random occurrences. For typical devices,
in order to obtain stable and ``smooth'' averages, we
need to consider an ensemble of about a hundred occurrences.

For the purpose of clarity, we have described the case
of a two-terminal devices. However, the method
can be applied without any variation for the computation of
many-terminal scattering matrices.

Using some algebra it is straightforward to verify the unitarity
of any $\tilde{S}$: adding the random phase term to the
scattering matrix of the $j$-th slice corresponds to substitute
the coherent matrix $S_j$ with $\tilde{S}_j = S_j \otimes
S_j^{\rm random}$, where $S_j^{\rm random}$ is a
scattering matrix in which the reflection matrices are zero,
and the transmission matrices are diagonal, their $n$-th
element ($n=1,\dots,N_{\rm mode}$) being $\exp(i\Delta \phi_{j,n})$.

It is easy to verify that $S_j^{\rm random}$ is unitary
by construction; since composition of unitary scattering
matrices provide a unitary scattering matrix, each $\tilde{S}$
is unitary.

The physical reason for unitarity of $\tilde{S}$ is the
conservation of the incoming current, whereas the time reversal
symmetry in the presence of a magnetic flux $\Phi$ implies the
validity of the Onsager-Casimir relations\cite{onsager} \be
\label{onsa} T_{pq}(\Phi)=T_{qp}(-\Phi)\; , \hspace{1cm}
R_{pp}(\Phi)=R_{pp}(-\Phi) \; , \ee where the labels $q$ and $p$
denote the leads of the system and $T_{pq}$ is the total
transmission probability from lead $p$ to lead $q$ (summed over
all modes), and $R_{pp}$ is the total reflection probability at
lead $p$.\cite{buttiker2} Once again, each $S_j^{\rm random}$ is
symmetric and independent of the magnetic field and therefore
obeys (\ref{onsa}); it is now sufficient to observe that a
composition of matrices obeying (\ref{onsa}) still provides a
matrix that obeys Onsager-Casimir relations.

Let us consider a traveling plane wave that loses phase
coherence as it propagates, but conserves its modulus.
One possible description of such situation is to
write the wave function as the sum of a coherent component
whose amplitude decays exponentially with propagation for a length $l$ as
$e^{-l/2l_\phi}$ and of an incoherent component
totally uncorrelated with the former, that ensures
conservation of the wave function modulus.
Another possible description is to add to the
phase of the traveling wave function after a length $l$ a random term,
with Gaussian distribution, zero average and standard deviation $\sigma$.

In order to derive the relation between $l_\phi$ and $\sigma$,
we consider the case of wave interference.
First, let us consider two coherent wave functions $\psi_1$ and $\psi_2$
of amplitude unity, obtained for example with a beam splitter.
We let them interfere again after both propagate along paths of length $l$.
In terms of the former description the amplitude of the interfering pattern is
\begin{equation}
|\psi_1+\psi_2|^2_{\max} - |\psi_1+\psi_2|^2_{\min} =4 \exp(-l/l_\phi).
\label{one}
\end{equation}

On the other hand, if we write the same two wave functions with the latter
description, they have amplitude unity and 
phases containing additional random terms
$\phi_{\text R}^1$ and $\phi_{\text R}^2$, respectively, 
that are uncorrelated, and
obey a Gaussian distribution with average zero and standard deviation $\sigma$.
The amplitude of the interfering pattern, in this case, is
\begin{equation}
\langle |e^{i \phi_{\text R}^1}+ e^{i \phi_{\text R}^2}|^2\rangle -
\langle |e^{i \phi_{\text R}^1}- e^{i \phi_{\text R}^2}|^2\rangle=
4\langle \cos(\phi_{\text R}^1-\phi_{\text R}^2) \rangle =
4 e^{-\sigma^2},
\label{two}
\end{equation}
where angle brackets mean statistical averaging,
and the last equation has been obtained using the fact that
$\phi_{\text R}^1-\phi_{\text R}^2$ is a Gaussian variable
of average zero and variance $2\sigma^2$, and
$\langle \cos \phi \rangle
=\int d\phi \, \cos \phi \, \exp [ -\phi^2/4\sigma^2 ]/\sqrt{4\pi\sigma^2}
=e^{-\sigma^2}$.

By comparing (\ref{one}) and (\ref{two}), we obtain $\sigma^2 = l/l_\phi$,
that, if we consider each single slice in which the structure is partitioned,
means
\begin{equation}\label{sigma}
    \sigma_j = \sqrt{\frac{x_{j+1}-x_j}{l_\phi}}.
\end{equation}


\section{Simulations}
\label{results}

In this section we use our proposed model for dephasing
to investigate the effect of decoherence on UCF and
Aharonov-Bohm oscillations in mesoscopic rings,
for which experimental results are available in the literature.

\begin{figure}[ht!]
\includegraphics[width=8cm]{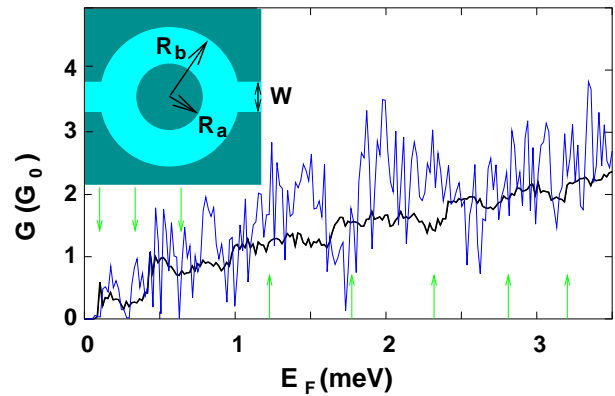}
\caption{(Color online).
Conductance of the AB ring as a function of the Fermi level
of electrons: Completely
coherent regime (thin solid line), partially incoherent regime corresponding
to $l_\phi=0.3 \mu$m (thick solid line).
Results are averaged on 100 random occurrences.
The arrows indicate the energies at which a new conducting channel
in the lead is opened.
Inset: the AB ring potential used in our simulations.
The internal radius $R_{\text a}$ is 350 nm,
whereas the external radius $R_{\text b}$ is 630 nm.
The width of the branches $W$ that connect the ring to drain and source
is 200 nm. $G_0$ is the conductance quantum $2e^2/h$.}
\label{ring}
\end{figure}

The Aharonov-Bohm ring (shown the inset of Fig.~\ref{ring})
is broadly used to perform phase coherence measurements
because it provides the possibility to obtain WL, UCF, as well as
pure interference effects.\cite{stone}
As far as this aspect is concerned,
analytical predictions of the dephasing rate 
are available,\cite{seelig} that agree very well
with the experiments.\cite{hansen} 

UCF appears when an external parameter that alters
the potential profile of the structure is varied.
Indeed, such conductance fluctuations are obtained
in experiments by varying the Fermi level $E_{\text{F}}$
of the electrons through the voltage on
the back gate or on a top gate.
The typical amplitude of conductance fluctuations
does not depend on the sample size or on the degree of disorder
and is of the order of the conductance quantum $2e^2/h$
in a purely coherent transport regime.\cite{altshuler,lee,beenakker}
If this is not the case, decoherence
smears out fluctuations restoring a staircase
when $G$ is plotted versus $E_{\text{F}}$ or versus the gate voltage, as
shown for example in the experiments of Ref.~\onlinecite{hansen,pedersen}.

We have simulated a symmetric AB ring structure defined by etching
on a GaAs/AlGaAs heterostructure. With reference to the inset of
Fig.~\ref{ring}, the internal radius of the ring is 350 nm, the
external radius is 630 nm, the width of the leads is 200 nm.
Bright regions correspond to a potential energy of 0 eV, 
dark regions to 0.2~eV.

The thin line in Fig.~\ref{ring} is the
computed conductance as a function of $E_{\text{F}}$
for fully coherent transport, while the thick line is the conductance
corresponding to a dephasing length $l_\phi=0.3$~$\mu$m. Results are obtained
by averaging on 100 random occurrences. The fluctuations clearly present
in the coherent regime are evidently smoothed out as decoherence
is introduced.

In addition, as a consequence of the introduction of decoherence
we are able to observe the loss 
of the universality in the conductance behavior,
that now presents a series of non-integer plateaus.
Such behavior is typically observed in experiments
\cite{picciotto,yacoby,hansen} as is due to backscattering
phenomena and to scattering at the interface between 1D and
2D electron gas. However, some degree of decoherence, 
which is always present in experiments,
is required to clearly reproduce the phenomenon with simulations.

\begin{figure}[ht!]
\includegraphics[width=8cm]{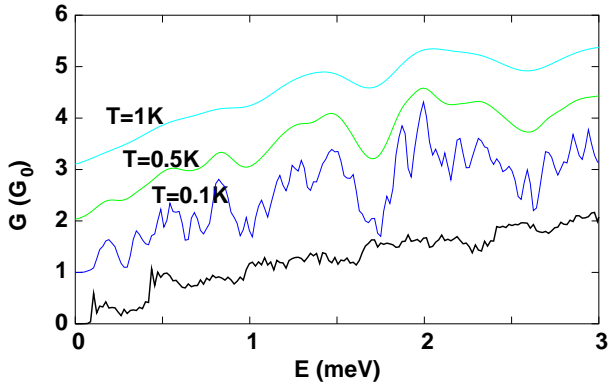}
\caption{(Color online).Comparison between the conductance of the AB ring
at 0 K computed with $l_\phi = 0.3$~$\mu$m
(thick line) and fully coherent conductances computed at finite temperature (thin lines).
For clarity of presentation each line is shifted by one unit of conductance }
\label{temp}
\end{figure}

We want to emphasize that dephasing introduced by
our model has a very different effect on the device
conductance than energy averaging due to
a finite temperature of the system.
In Fig.~\ref{temp} we plot the conductance
of the same ring in Fig.~\ref{ring} versus $E_{\text{F}}$
and compare it with
the thermal-averaged conductance computed in the case of
fully coherent transport:
\be \nonumber
\overline{G}(T)=-\int dE \, G(E) \, \frac{\partial f(E,T)}{\partial E} \; ,
\ee
where $f(E,T)$ is the Fermi-Dirac occupation factor.
As can be seen, the coherent conductance at different temperatures
does not exhibit the non-integer plateaus previously observed and
has a behavior qualitatively different from partially coherent conductance.

\begin{figure}[ht!]
\includegraphics[width=8cm]{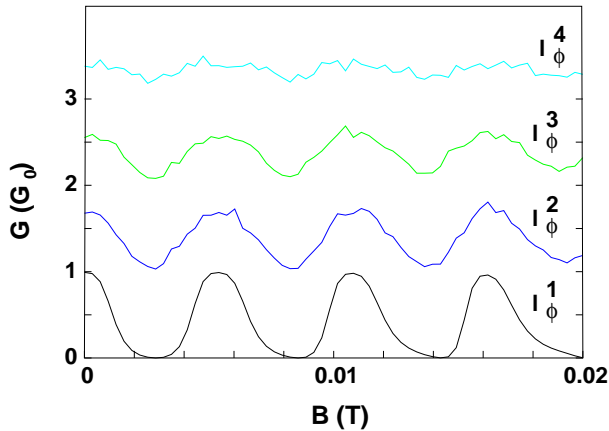}
\caption{(Color online).
The conductance oscillations of the AB ring shown in the inset
of Fig.~\ref{ring} as a function
of the magnetic field, for dephasing length
$l^1_{\phi}=\infty, l^2_{\phi}=5 \,\mu {\rm m}, l^3_{\phi}=1
\, \mu {\rm m}, l^4_{\phi}=0.3 \, \mu {\rm m}$.
$G$ is averaged over 100 simulation runs.
For clarity of presentation, each line is shifted by one conductance unit.
}
\label{mod}
\end{figure}

Another way to verify how dephasing influences the transport properties
of the ring structure is represented by the study of magnetoconductance.
In our code we have added the effects of an external magnetic field
${\bf B}=(0,0,B)$ perpendicular to the propagation plane $xy$.
We adopt the transverse gauge
${\bf A}=(-By,0,0)$ for the vector potential ${\bf A}$
as described in Ref.~\onlinecite{governale}.
Due to the AB ring geometry the phase difference of
wave functions propagating along the two branches depends
on the magnetic field as
$\int({\bf p}-e {\bf A})/h \cdot d{\bf r}$,
generating the magnetoconductance oscillations.
The oscillation period can be equal to the quantum flux $h/e$
or to the submultiples $h/ne$ when coherent backscattering is present
and the electron turns around the ring more times.
As expected, decoherence suppresses the amplitude of
magnetoconductance. Results for the AB ring geometry of the
inset of Fig.~\ref{ring} are shown in Fig.~\ref{mod}, where it
is possible to appreciate the transition from a coherent transport regime
to an only partially coherent one as $l_\phi$ is decreased.

\begin{figure}[ht!]
\includegraphics[width=8cm]{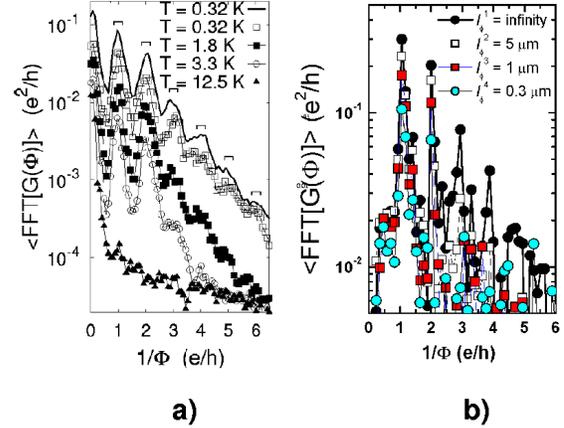}
\caption{(Color online).
Left:
Fast Fourier Transform (FFT) of the AB experimental oscillations
measured by Hansen et al. (Fig. 2 of Ref.~\onlinecite{hansen})
at different temperatures.
Right: the same FFT obtained with our simulations for values
of the dephasing length $l^1_{\phi}=\infty$,
$l^2_{\phi}=5$~$\mu$m, $l^3_{\phi}=1$~$\mu$m,
$l^4_{\phi}=0.3$~$\mu$m. In the simulation results, the DC
component has been removed.}
\label{osc}
\end{figure}

In Fig.~\ref{osc} we show a comparison with experimental
results presented by Hansen ${\it et \, al.}$ \cite{hansen}
for a symmetric AB ring with internal radius 280 nm,
external radius 560 nm, and wire width 100 nm.
On the left we show the Fast Fourier Transform (FFT)
of experimental magnetoconductance oscillations
(Fig.~2 of Ref.~\onlinecite{hansen})
for different values of temperature, while on the right
we show the same quantity computed with our model
for different values of the dephasing length.
In both cases all frequency components are damped
by decoherence. At small values of $l_\phi$ only
the first peak corresponding to the $h/e$
frequency is clearly visible.

\begin{figure}[ht!]
\includegraphics[width=8cm]{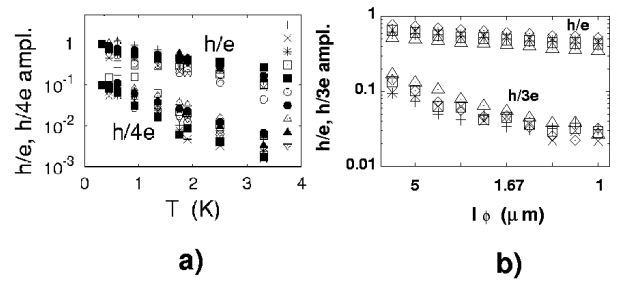}
\caption{
Left:
Measured oscillation amplitude for the
$h/e$ and $h/4e$ frequencies (Fig.~2 of Ref.~\onlinecite{hansen}).
Right:
Amplitude of the $h/e$ and $h/3e$ oscillations
plotted as a function of the dephasing length $l_\phi$.
Each dot corresponds to a different Fermi energy of the propagating electrons.
}
\label{dep}
\end{figure}

It is important to verify whether the dephasing
length microscopically introduced by our model
through Eq.~(\ref{sigma}) agrees with the value that can be extracted
from the electrical properties of the device.
Following Ref.~\onlinecite{hansen} we assume
that the amplitude $A_{n}$ of the $h/ne$ oscillation
can be written as
\be
\label{expo}
A_n \propto e^{-nL/l_\phi} \; ,
\ee
where $L$ is the circumference of the ring.
Such assumption is confirmed by the experimental results
shown in Fig.~\ref{dep}a (Fig.~2 of Ref. \onlinecite{hansen}).

The FFT of the simulated oscillation amplitudes for different $n$
exhibit an exponential dependence on $l_\phi$
as shown in the semilog plot
of Fig.~\ref{dep}b for the cases $n$=1 and $n$=3.
Both the slopes of the $h/3e$ and the $h/e$ oscillations,
according to Eq.~(\ref{expo}), are
consistent with the nominal value provided by Eq.~(\ref{sigma}).
Results for $n>3$ are not reliable in the whole range of $l_\phi$
due to numerical fluctuations and therefore are not shown.

\begin{figure}[ht!]
\includegraphics[width=8cm]{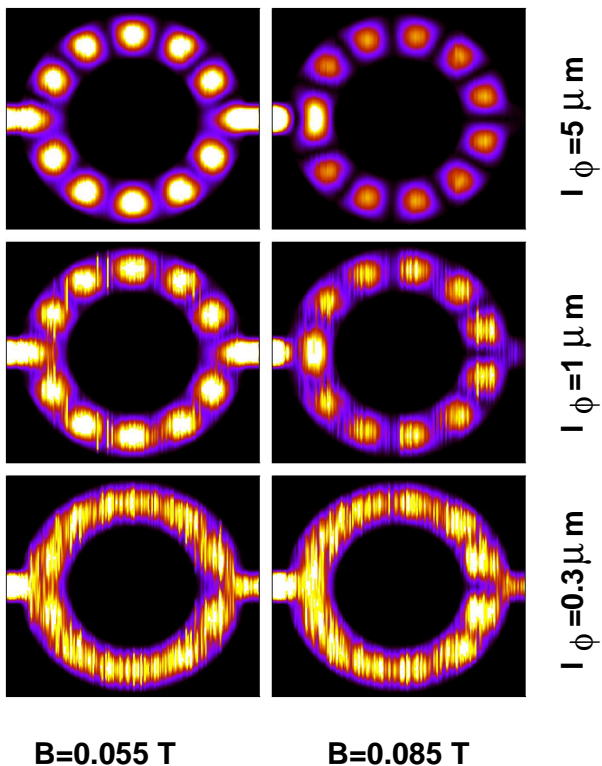}
\caption{(Color online).
Partial density of states
for different dephasing lengths in the AB ring of Fig.~1
when only the first mode is populated.
The magnetic field $B$ is 55 mT and 85 mT, corresponding
to a maximum and a minimum of conductance (see Fig.~\ref{mod}).}
\label{dos}
\end{figure}

Finally, we emphasize the possibility of a microscopic
description of the effects of decoherence in the system.
In Fig.~\ref{dos} we show the partial local density of
states corresponding to one mode injected from
the left $\rho(x,y,E) \propto |\psi(x,y,E)|^2$
for three different values of $l_\phi$ and
two different values of the magnetic field $B$, corresponding
to the cases of maximum constructive interference ($B = 55$~mT)
and maximum destructive interference ($B = 85$~mT),
as can be seen in Fig.~3. Also in this case the density
of states is obtained by averaging over 100 Monte Carlo runs.

For $l_\phi = 5$~$\mu$m transport is almost fully coherent,
a clear pattern of nodes forms in both branches, and in the
output lead we have maximum modulation of the density of states
as a function of $B$. For smaller dephasing lengths the
stationary wave pattern in the branches smooths out. In
particular, for $l_\phi = 0.3$~$\mu$m, when the interference
pattern is almost destroyed, as can be seen in Fig.~3,
the density of states is quasi-constant in the branches and
in the leads.


\section{Conclusions}

In this paper we have presented a phenomenological
microscopic model for the simulation of dephasing
in mesoscopic devices. The stochastic nature of
the dephasing process is taken into account with
a Monte Carlo methodology used to extract
average conductance, magnetoconductance and density of states.
We have shown that the proposed method ensures the
physical validity of each occurrence of the scattering matrix.

Here we want to underline the fact that the proposed
method provides a unique description applicable to systems
with an arbitrary degree of dephasing. This represents the main
advantage of the proposed method, since common methods for
determining transport properties of generic devices
consider only the
limit of completely coherent transport (with
scattering matrix or recursive Green's functions
techniques) or the limit of fully incoherent
transport (with semiclassical approaches).

We have also shown that such method allows to recover
experimental results observed in Aharonov-Bohm rings, when
a certain degree of decoherence is always present
and responsible for some typical features,
such as the non-integer conductance plateaus at zero
magnetic field.

We believe that the proposed model can be very
useful in understanding the effect of dephasing
on the transport properties of mesoscopic devices,
and enables to accurately reproduce
experimental results with numerical simulations. It can also
have a significant effect in assessing the effect
of dephasing on the noise properties of nanoscale devices.

\section{Acknowledgments}

The authors would like to thank M. Governale for interesting discussions.
The work presented in this paper has been supported by
the IST NANOTCAD project (EU contract IST-1999-10828),
and by the SINANO Network of Excellence (EU contract 506844).



\end{document}